\documentclass[aps,prd,twocolumn,showpacs,floats,floatfix,letterpaper,nofootinbib,superscriptaddress]{revtex4}
\usepackage{amssymb,amsmath,latexsym,mathrsfs}
\usepackage{graphicx}
\usepackage{graphicx,subfigure}
\usepackage{epsfig}
\usepackage{multirow}
\usepackage{array} 
\usepackage{varioref,xr-hyper}

\newcommand{\neff}{N_{\textrm{eff}}}

\newcommand{\cvis}{c_{\rm vis}^2}
\newcommand{\ceff}{c_{\rm eff}^2}

\begin{document}

%%%%%%%%%%%%%%%%%%%%%%%%%%%%%%%%%%%%%%%%
%%%%%%%%%%%%%%%%%%%%%%%%%%%%%%%%%%%%%%%%

\title{Dark Radiation and interacting scenarios}

\author{Roberta Diamanti}
\affiliation{Department of Physics, Universit\`a Roma Tre, Via della Vasca Navale 84, 00146, Rome, Italy }
\author{Elena Giusarma}
\affiliation{IFIC, Universidad de Valencia-CSIC, 46071, Valencia, Spain}
\author{Olga Mena}
\affiliation{IFIC, Universidad de Valencia-CSIC, 46071, Valencia, Spain}
\author{Maria Archidiacono}
\affiliation{Department of Physics and Astronomy
University of Aarhus, DK-8000 Aarhus C, Denmark}
\author{Alessandro Melchiorri}
\affiliation{Physics Department and INFN, Universita' di Roma 
	``La Sapienza'', Ple.\ Aldo Moro 2, 00185, Rome, Italy}

\begin{abstract}
An extra dark radiation component can be present in the universe in the form of sterile neutrinos, axions or other very light degrees of freedom which may interact with the dark matter sector. We derive here the cosmological constraints on the dark radiation abundance, on its effective velocity and on its viscosity parameter from current data in dark radiation-dark matter coupled models. The cosmological bounds on the number of extra dark radiation species do not change significantly  when considering interacting schemes. We also find that the constraints on the dark radiation effective velocity are degraded by an order of magnitude while the errors on the viscosity parameter are a factor of two larger when considering interacting scenarios. If future Cosmic Microwave Background data are analysed assuming a non interacting model but the dark radiation and the dark matter sectors interact in nature, the reconstructed values for the effective velocity and for the viscosity parameter will be shifted from their standard $1/3$ expectation, namely $\ceff=0.34^{+0.006}_{-0.003}$ and $\cvis=0.29^{+0.002}_{-0.001}$ at $95\%$~CL for the future COrE mission data.
\end{abstract}


\maketitle

%%%%%%%%%%%%%%%%%%%%%%%%%%%%%%%%%%%%%%%%
\section{Introduction}

From observations of the Cosmic Microwave Background (CMB) and large scale structure (LSS) we can probe the fundamental properties of the constituents of the cosmic dark radiation background. The energy density of the total radiation component reads
\begin{equation}
 \rho_{rad} = \left[1 + \frac{7}{8} \left(\frac{4}{11}\right)^{4/3}\neff\right]\rho_{\gamma} \, ,
\end{equation}
where $\rho_{\gamma}$ is the current energy density of the CMB and $\neff$ is a free parameter, defined as the effective number of relativistic degrees of freedom in the cosmic dark radiation background. In the standard scenario, the expected value is $\neff=3.046$, corresponding to the three active neutrino contribution and considering effects related to non-instantaneous neutrino decoupling and QED finite temperature corrections to the plasma. The most recent CMB data analyses gives $\neff = 3.89\pm 0.67$ ($68\%$~CL)~\cite{Hinshaw:2012fq}, see also Refs.~\cite{Giusarma:2012ph,Mangano:2006ur,Hamann:2007pi,Reid:2009nq,Komatsu:2010fb,Izotov:2010ca,Hamann:2010pw,Hou:2011ec,Keisler:2011aw,Smith:2011es,Dunkley:2010ge,darkr,Hamann:2011hu,Nollett:2011aa,Smith:2011ab,darkr2,concha}. The simplest scenario to explain the extra dark radiation $\Delta\neff\equiv\neff-3.046$ arising from cosmological data analyses assumes the existence of extra sterile neutrino species. However, there are other possibilities which are as well closely related to minimal extensions to the standard model of elementary particles, as axions, extra dimensions or asymmetric dark matter models.

Dark radiation, apart from being parametrized by its effective number of relativistic degrees of freedom, $\neff$, is also characterized by its clustering properties, i.e, its rest-frame speed of sound, $\ceff$, and its viscosity parameter, $\cvis$, which controls the relationship between velocity/metric shear and anisotropic stress in the dark radiation background~\cite{Hu:1998kj}. A value of $\cvis$ different from zero, as expected in the standard scenario, sustains the existence of dark radiation anisotropies~\cite{Hu:1998tk}. The standard value of $\cvis = 1/3$ implies that the anisotropies in the dark radiation background are present and they are identical to the neutrino viscosity. On the other hand, the case $\cvis = 0$ cuts the Boltzmann hierarchy of the dark radiation perturbations at the quadrupole, representing a perfect fluid with density and velocity (pressure) perturbations exclusively. A value of $\ceff$ different from the canonical $\ceff = 1/3$ leads to a non-adiabatic dark radiation pressure perturbation, i.e. $(\delta p - \delta \rho/3)/\bar{\rho} = (\ceff - 1/3)\delta_{dr}^{rest}$, where $\delta_{dr}^{rest}$ is the density perturbation in the rest frame, where the dark radiation velocity perturbation is zero.

Interacting dark radiation arises naturally in the so-called asymmetric dark matter models (see e.g.~\cite{asy} and references therein), in which the dark matter production mechanism is similar and related to the one in the baryonic sector. In these models, there exists a particle-antiparticle asymmetry at high temperatures in the dark matter sector. The thermally symmetric dark matter component will annihilate and decay into dark radiation degrees of freedom. Since the dark radiation and the dark matter fluids are interacting, there was an epoch in the early Universe in which these two dark fluids were strongly coupled. This results in a tightly coupled fluid with a pressure producing oscillations in the matter power spectrum analogous to the acoustic oscillations in the baryon-photon fluid before the recombination era. Due to the presence of a dark radiation-dark matter interaction, the clustering properties of the dark radiation component can be modified~\cite{Smith:2011es}. In other words, if dark radiation is made of interacting particles, the values of the clustering parameters $\ceff$ and $\cvis$ may differ from the canonical $\ceff=\cvis = 1/3$.

The cosmological implications of interacting dark radiation with canonical clustering properties have been carefully explored in Refs.~\cite{Mangano:2006mp,Serra:2009uu,asyus}, see also the recent work of Ref.~\cite{concha}. Here we generalize the analysis and leave the three dark radiation parameters $\Delta \neff$, $\ceff$ and $\cvis$ to vary freely within a $\Lambda$CDM scenario with a dark radiation-dark matter interaction. We will see that the bounds from current cosmological data on the dark radiation properties derived in non interacting schemes in Refs.~\cite{darkr,Smith:2011es,darkr2} will be, in general, relaxed, when an interaction between the dark radiation and the dark matter fluids is switched on. While the bounds on the number of extra dark radiation species will not be largely modified in coupled schemes, the errors on the dark radiation effective velocity and viscosity parameters will be drastically increased in interacting scenarios. We also show here how future CMB measurements, as those from the Planck~\cite{planck} and COrE~\cite{core} experiments, can lead to large biases on the dark radiation clustering parameters if the dark radiation and dark matter fluids interact in nature but the data is analyzed in the absence of such a coupling.

The paper is organized as follows. Section~\ref{sec:DRM} presents the parametrization used for dark radiation, describing the dark radiation-dark matter interactions explored here and their impact on the cosmological observables used in the analysis, as the CMB temperature anisotropies and the matter power spectrum. In Sec.~\ref{sec:data} we describe the data sets used in the Monte Carlo analyses presented in Sec.~\ref{sec:results}, which summarizes the constraints on interacting dark radiation properties from current cosmological data. Future CMB dark radiation measurements are presented in Sec.~\ref{sec:forecasts}. We draw our conclusions in Sec.~\ref{sec:conclusions}.

\section{Dark radiation-dark matter interaction model} 
\label{sec:DRM}
The evolution of the dark radiation linear perturbations reads~\cite{Hu:1998kj} 

\begin{align}
& \dot{\delta}_{dr} - \frac{\dot{a}}{a}(1 - 3\ceff)\left(\delta_{dr} + 4\frac{\dot{a}}{a}\frac{\theta_{dr}}{k^2}\right) + \frac{4}{3}\theta_{dr} + \frac{2}{3}\dot{h} =0~; \\  \label{NU} 
& \dot{\theta}_{dr} - 3k^2 \ceff\left(\frac{1}{4}\delta_{dr} + \frac{\dot{a}}{a}\frac{\theta_{dr}}{k^2}\right) + \frac{\dot{a}}{a}\theta_{dr} + \frac{1}{2}k^2\pi_{dr}=0~; \\ \label{q}
 &\dot{\pi}_{dr} + \frac{3}{5}kF_{dr,3} - \frac{8}{5}\cvis\left(\frac{\theta_{dr}}{k} + \sigma \right) = 0~; \\
 &\frac{2l + 1}{k}\dot{F}_{dr,3} - lF_{dr,l-1} + (l + 1)F_{dr,l+1} = 0 \, \quad l \geq 3~, 
\end{align}
\noindent where the dots refer to derivatives with respect to conformal time, $a$ is the scale factor, $k$ is the wavenumber, $\ceff$ is the effective sound speed, $\cvis$ is the viscosity parameter, $\delta_{dr}$ and $\theta_{dr}$ are the dark radiation energy density perturbation and velocity divergence, respectively, and $F_{dr,l}$ are the higher order moments of the dark radiation distribution function. In the set of equations above, $\pi_{dr}$ is the anisotropic stress perturbation, and $\sigma$ is the metric shear, defined as $\sigma = (\dot{h}+6\dot{\eta})/(2k)$, with $h$ and $\eta$ the scalar metric perturbations in the synchronous gauge. The anisotropic stress would affect the density perturbations, as in the case of a real fluid, in which the stress represents the viscosity, damping the density perturbations. The relationship between the metric shear and the anisotropic stress can be parametrized through a ``viscosity parameter'', $\cvis$~\cite{Hu:1998kj}:

\begin{equation}\label{SHEAR-STRESS}
 \dot{\pi} = -3\frac{\dot{a}}{a}\pi + 4\cvis\left(\theta + \sigma \right) \, ,
\end{equation}

\noindent where $\theta$ is the divergence of the fluid velocity. Although the perturbed Einstein and energy-momentum conservation equations are enough to describe the evolution of the cosmological perturbations of non-relativistic particles, it is convenient to introduce the full distribution function in phase space to follow the perturbation evolution of relativistic particles, that is, to consider their Boltzmann equation. In order to determine the evolution equation of dark radiation, the Boltzmann equation is transformed into an infinite hierarchy of moment equations, that must be truncated at some maximum multipole order $l_{max}$. Then, the higher order moments of the distribution function are truncated with appropriate boundary conditions, following Ref.~\cite{Lewis:1999bs}.

In the presence of a dark radiation-dark matter interaction, the Euler equations for these two dark fluids read

\begin{eqnarray}
 \dot{\theta}_{dm} &=& -\frac{\dot{a}}{a}\theta_{dm} + \frac{4\rho_{dr}}{3\rho_{dm}}an_{dm}\sigma_{dm-dr}(\theta_{dr} - \theta_{dm})~, \label{eq:CDM}\\
 \dot{\theta}_{dr} &=& \frac{1}{4}k^2\left(\delta_{dr} - 2 \pi_{dr}\right) + an_{dm}\sigma_{dm-dr}(\theta_{dm} - \theta_{dr})  \label{eq:NEUTRINOS}
\end{eqnarray}

\noindent where the momentum transfer to the dark radiation component is given by $an_{dm}\sigma_{dm-dr}(\theta_{dm} - \theta_{dr})$. Indeed, the former quantity is the differential opacity, which gives the scattering rate of dark radiation by dark matter~\cite{Mangano:2006mp,Serra:2009uu}. The complete Euler equation for dark radiation, including the interaction term with the dark matter fluid, reads

\begin{eqnarray}
 \dot{\theta}_{dr}& =& 3k^2 \ceff\left(\frac{1}{4}\delta_{dr} + \frac{\dot{a}}{a}\frac{\theta_{dr}}{k^2}\right) - \frac{\dot{a}}{a}\theta_{dr} - \frac{1}{2}k^2\pi_{dr}\nonumber \\ 
&+& an_{dm}\sigma_{dm-dr}(\theta_{dm} - \theta_{dr})~.
\label{1-INT}
\end{eqnarray}

\noindent Following Refs.~\cite{Mangano:2006mp,Serra:2009uu} we parameterize the coupling between massless neutrinos and dark matter through a cross section given by
\begin{equation}
 \langle\sigma_{dm-dr}|v|\rangle \sim Q_0\, m_{dm}~,
\end{equation}

\noindent if it is constant, or
\begin{equation}
 \langle\sigma_{dm-dr}|v|\rangle \sim \frac{Q_2}{a^2}\, m_{dm}~,
\end{equation}

\noindent if it is proportional to $T^2$, where the parameters $Q_0$ and $Q_2$ are constants in cm$^2$ MeV$^{-1}$ units.  
It has been shown in Ref.~\cite{asyus} that the cosmological implications of both constant and T-dependent interacting cross sections are very similar. Therefore, in the following, we focus on the constant cross section case, parameterized via $Q_0$. 

\begin{figure*}[!htb]
\begin{tabular}{c}
\includegraphics*[width=7.cm,angle=270]{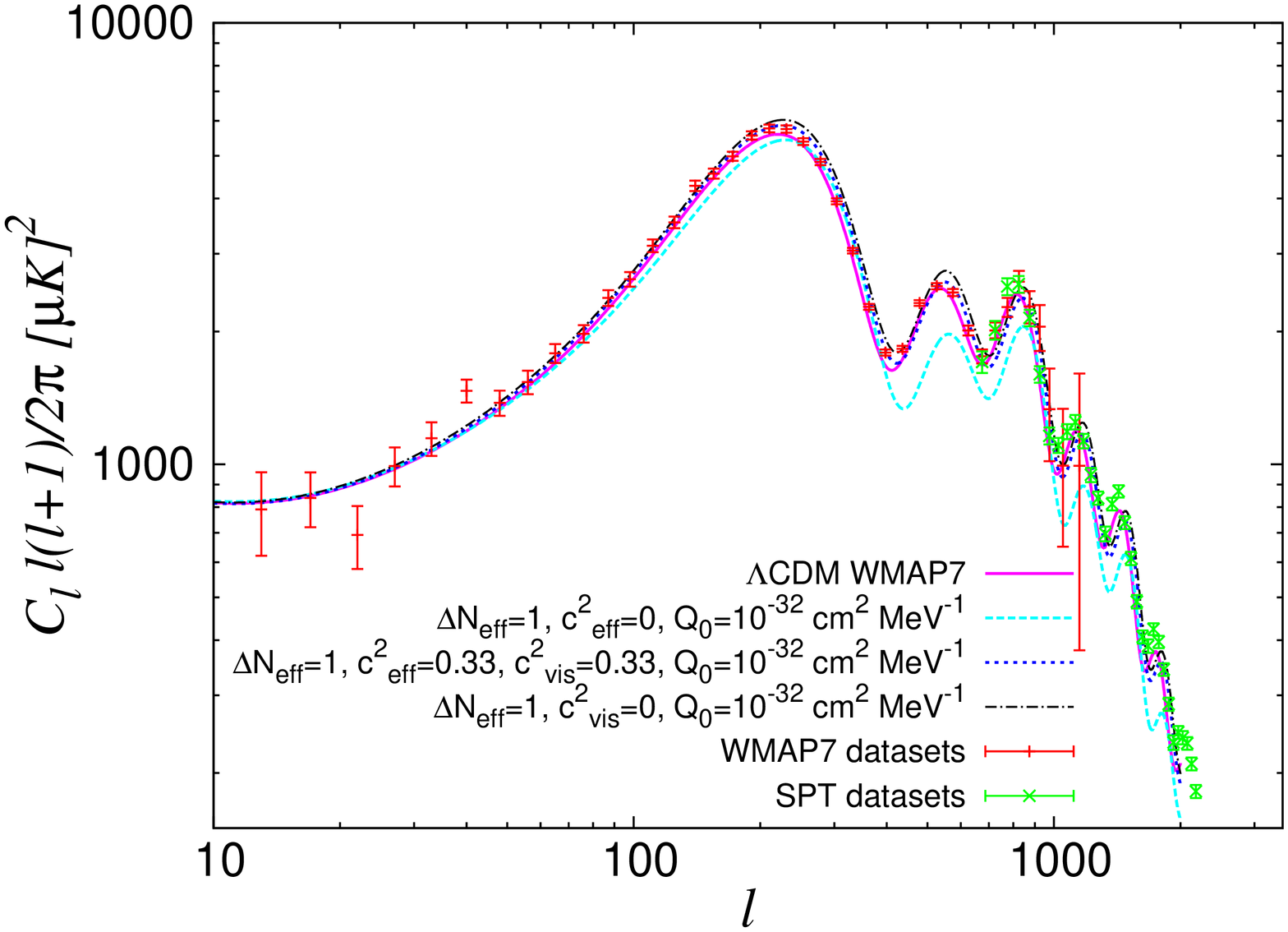}\\
\includegraphics*[width=7.cm,angle=270]{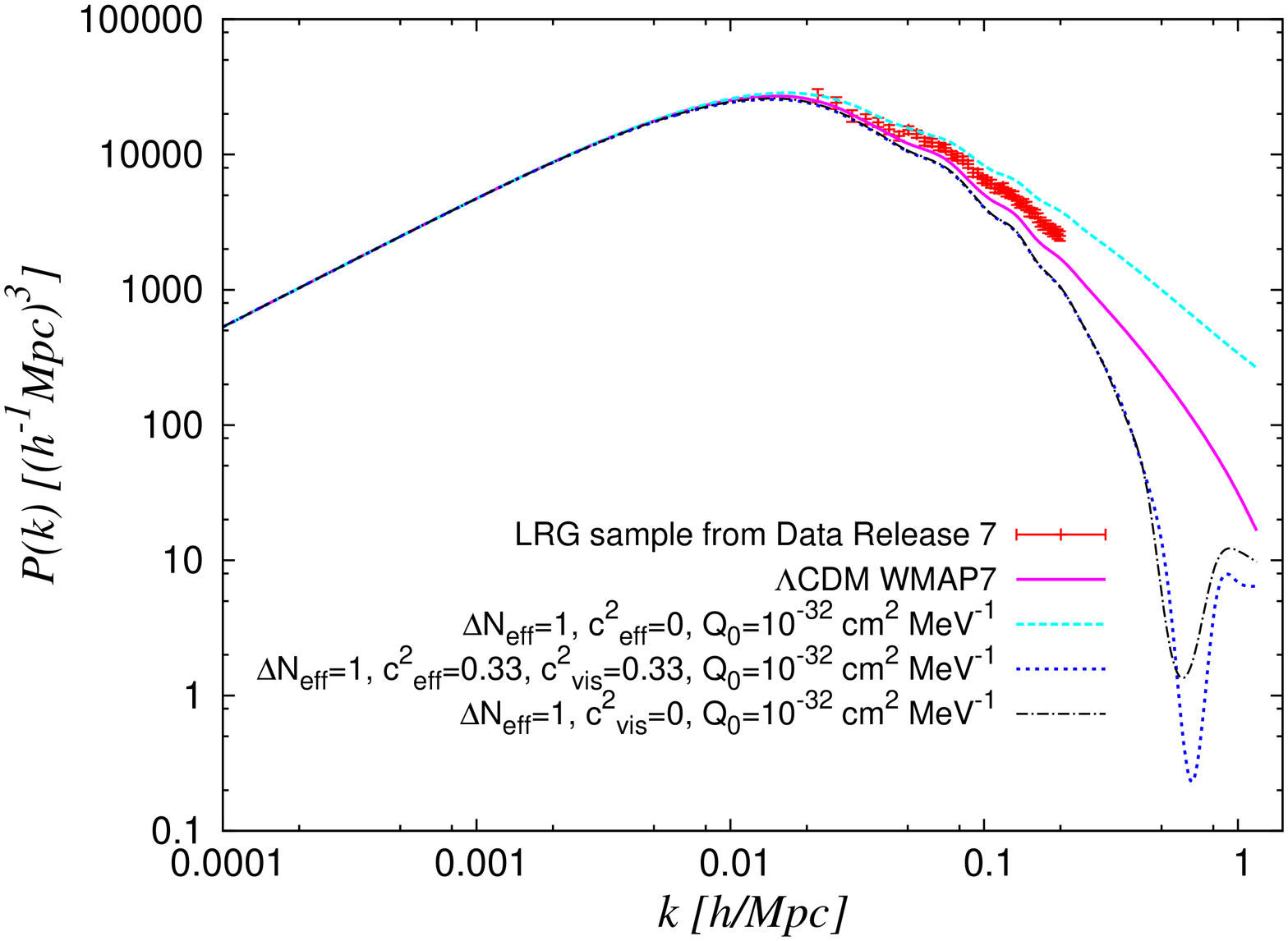}\\
\end{tabular}
 \caption{\small{Upper panel: The magenta lines depict the CMB temperature power spectra $C_l^{TT}$ for the best fit parameters for a $\Lambda$CDM model from the WMAP seven year data set. The dotted curve shows the scenario with a constant interacting cross section with $Q_0 = 10^{-32}$cm$^2$ MeV$^{-1}$ for $\Delta\neff = 1$ and assuming canonical values for $\ceff=\cvis=1/3$. The dashed (dot dashed) curve illustrates the same interacting scenario but with $\ceff=0$ and $\cvis=1/3$ ($\ceff=1/3$ and $\cvis=0$). We depict as well the data from the WMAP and SPT experiments, see text for details. Lower panel: matter power spectrum for the different models described in the upper panel. The data correspond to the clustering measurements of luminous red galaxies from SDSS II DR7~\cite{beth}.}}
\label{WS}
\end{figure*}

Figure \ref{WS}, upper panel, shows the CMB temperature anisotropies for $Q_0 = 10^{-32}$ cm$^2$\,MeV$^{-1}$ and one dark radiation interacting species, i.e. $\Delta\neff=1$, as well as for the non interacting case for the best fit parameter values from WMAP seven year data analysis~\cite{Komatsu:2010fb,wmap7} together with WMAP and South Pole Telescope (SPT) data~\cite{Keisler:2011aw}. We illustrate the behavior of the temperature anisotropies for different assumptions of the dark radiation clustering parameters. Notice that the presence of a dark radiation-dark matter interaction enhances the height of the CMB peaks due to both the presence of an extra radiation component ($\Delta\neff$) and the fact that dark matter is no longer pressureless (due to a non zero $Q_0$). Therefore $\Delta\neff$ and $Q_0$ will be negatively correlated.
The location of the peaks also changes, mostly due to the presence of extra radiation $\Delta\neff$. The peaks will be shifted to higher multipoles $\ell$ due to changes in the acoustic scale, given by
\begin{equation}
\theta_A=\frac{r_s(z_{rec})}{r_\theta(z_{rec})}~,
\end{equation}
where $r_\theta (z_{rec})$ and $r_s(z_{rec})$ are the comoving angular diameter distance to the last scattering surface and the sound horizon at the recombination epoch $z_{rec}$, respectively. Although $r_\theta (z_{rec})$ almost remains the same for different values of $\Delta\neff$, $r_s(z_{rec})$ becomes smaller when $\Delta\neff$ is increased. Therefore, the positions of the acoustic peaks are shifted to higher multipoles (smaller angular scales) if the value of $\Delta \neff$ is increased. Notice, however, that this effect can be compensated by changing the cold dark matter density, in such a way that $z_{rec}$ remains fixed, see Ref.~\cite{Hou:2011ec}.
Changing $\cvis$ modifies the ability of the dark radiation to free-stream out of the potential wells~\cite{Hu:1995fqa,Hu:1995en,Bowen:2001in}. Notice from Fig.~\ref{WS} (upper panel), that lowering $\cvis$ to the value $\cvis = 0$, the TT power spectrum is enhanced with respect to the standard case without the dark radiation and the dark matter species interacting. This situation can be explained, roughly, as the dark radiation component becoming a perfect fluid. That is, we are dealing with a single fluid characterized by an effective viscosity. Disregarding the fluid nature and the physical origin of the viscosity, the general consideration holds: for a given perturbation induced in the fluid, the amplitude of the oscillations that the viscosity produces (see, e.g. \cite{Smith:2011es}) increases as the viscosity is reduced. Therefore, lowering $\cvis$ diminishes the amount of damping induced by the dark radiation viscosity, and, consequently, in this case, the amplitude of the CMB oscillations will increase, increasing in turn the amplitude of the angular power spectrum. Therefore, we expect the interaction strength size $Q_0$ and the $\cvis$ parameter to be positively correlated. 

On the other hand, a change of $\ceff$ implies a decrease of pressure perturbations for the dark radiation component in its rest frame. As shown in Fig.~\ref{WS} (upper panel), a decrease in $\ceff$ from its canonical $1/3$ to the value $\ceff = 0$ leads to a damping of the CMB peaks, since dark radiation is behaving as a pressureless fluid from the perturbation perspective. In the case of $\ceff$, we expect this parameter to be negatively correlated with $Q_0$.

Figure~\ref{WS} (lower panel) depicts the matter power spectrum in the presence of a dark radiation-dark matter interaction for different values of the dark radiation clustering parameters (including the standard case with $\ceff=\cvis=1/3$) for one dark radiation interacting species, i.e. $\Delta\neff = 1$. We illustrate as well the case of a pure $\Lambda$CDM universe. Notice that, since the dark matter fluid is interacting with the dark radiation component, the dark matter component is no longer presureless, showing damped oscillations. The smaller wave mode at which the interaction between the dark fluids will leave a signature on the matter power spectrum is roughly $k_f\sim a_f H(a_f)$, which corresponds to the size of the universe at the time that the dark radiation-dark matter interaction becomes ineffective~\cite{Mangano:2006mp,Serra:2009uu,asyus}, i.e. when $H(a_f)=\Gamma(a_f)$ (being $H$ the Hubble parameter and $\Gamma$ the effective dark radiation-dark matter scattering rate $\frac{\rho_{dr}}{\rho_{dm}}n_{dm}\langle\sigma_{dm-dr}|v|\rangle$). For the case of constant dark radiation-dark matter interacting cross section, the typical scale $k_f$ reads, for $\Delta\neff=1$:

\begin{equation}
\label{eq:scale1}
k_f\sim 0.7 \left(\frac{10^{-32}~{\rm cm}^2~{\rm MeV}^{-1}}{Q_0}\right)^{1/2}~h\textrm{Mpc}^{-1}~, 
\end{equation}

Notice however from Fig.~\ref{WS} (lower panel) that, while varying $\cvis$ the matter power spectrum barely changes, a change in $\ceff$ changes dramatically the matter power spectrum, washing out any interacting signature. For instance, if $\ceff=0$, dark radiation is a presureless fluid which behaves as dark matter, inducing an enhancement of the matter fluctuations, and, consequently, the presence of a dark radiation-dark matter interaction will not modify the matter power spectrum, see the lower panel of Fig.~\ref{WS}. Therefore, one might expect a degeneracy between the dark radiation-dark matter coupling and the dark radiation $\ceff$ parameter: the larger the interaction is, the smaller $\ceff$ should be to compensate the suppression of power at scales $k\sim k_f$.

\section{Data}
\label{sec:data}

In order to constrain the dark radiation parameters $\Delta\neff$, $\ceff$ and $\cvis$, as well as the size of the dark radiation-dark matter interaction, we have modified the Boltzmann CAMB code \cite{Lewis:1999bs} including the dark radiation-dark matter interaction scenario. Then, we perform a Monte Carlo Markov Chain (MCMC) analysis based on the publicly available MCMC package \texttt{cosmomc} \cite{Lewis:2002ah}. We consider a $\Lambda$CDM cosmology with $\Delta\neff$ dark radiation species interacting with the dark matter and three massless active neutrinos. This scenario is described by the following set of parameters:
\begin{equation*}
  \{\omega_b,\: \omega_c,\: \Theta_s,\: \tau, \: n_s,\: \log[10^{10}A_{s}],\: \Delta\neff, \: \cvis, \: \ceff, \: Q_0\},
\end{equation*}
where $\omega_b\equiv\Omega_bh^{2}$ and $\omega_c\equiv\Omega_ch^{2}$ are the physical baryon and cold dark matter energy densities,
$\Theta_{s}$ is the ratio between the sound horizon and the angular diameter distance at decoupling, $\tau$ is the optical depth,
$n_s$ is the scalar spectral index, $A_{s}$ is the amplitude of the primordial spectrum, $\Delta\neff$ is the extra dark radiation component, $\cvis$ is the viscosity parameter, $\ceff$ is the effective sound speed and $Q_0$, in units of cm$^2$ MeV$^{-1}$, encodes the dark radiation-dark matter interaction. The flat priors considered on the different cosmological parameters are specified in Tab.~\ref{tab:priors}.

\begin{table}[h!]
\begin{center}
\begin{tabular}{c|c}
\hline\hline
 Parameter & Prior\\
\hline
$\Omega_{b}h^2$ & $0.005 \to 0.1$\\
$\Omega_{c}h^2$ & $0.01 \to 0.99$\\
$\Theta_s$ & $0.5 \to 10$\\
$\tau$ & $0.01 \to 0.8$\\
$n_{s}$ & $0.5 \to 1.5$\\
$\ln{(10^{10} A_{s})}$ & $2.7 \to 4$\\
$\Delta\neff$ &  $0  \to 10$\\
$\cvis$ &  $0 \to 1$\\
$\ceff$ &  $0 \to 1$\\
$\log_{10}\left(Q_0/10^{-32}\textrm{cm}^2 \textrm{MeV}^{-1} \right)$ & $-4\to 0$\\
\hline\hline
\end{tabular}
\caption{\small{Uniform priors for the cosmological parameters considered here.}}
\label{tab:priors}
\end{center}
\end{table}

For CMB data, we use the seven year WMAP data \cite{Komatsu:2010fb,wmap7}  (temperature and polarization) by means of the likelihood supplied by the WMAP collaboration. We consider as well CMB temperature anisotropies from the SPT experiment~\cite{Keisler:2011aw}, which provides highly accurate measurements on scales $\lesssim 10$ arcmin. We account as well for foreground contributions, adding the SZ amplitude $A_{SZ}$, the amplitude of the clustered point source contribution, $A_C$, and the amplitude of the Poisson distributed point source contribution, $A_P$, as nuisance parameters in the CMB data analysis.

Furthermore, we include the latest constraint from the Hubble Space Telescope (HST)~\cite{Riess:2011yx} on the Hubble parameter $H_0$. Separately, we also add Supernovae Ia luminosity distance data from the 3 year Supernova Legacy Survey (SNLS3)~\cite{snls3}, adding in the MCMC analysis two extra nuisance parameters, which are related to the intrinsic supernova magnitude dependence on stretch (which measures the shape of the SN light curve) and color, see Ref.~\cite{snls3} for details. We do not consider here the addition of HST and SNLS3 measurements simultaneously because these two data sets are not independent.

Galaxy clustering measurements are also added in our analysis via BAO data from the CMASS sample in Data Release 9~\cite{DR9} of the Baryon Oscillation Spectroscopic Survey (BOSS)~\cite{boss,boss2012}, with a median redshift of $z=0.57$~\cite{anderson}, as well as from the LRG sample from Data Release 7 with a median redshift of $z=0.35$~\cite{nikhil}, and from the 6dF Galaxy Survey 6dFGS at a lower redshift $z=0.106$~\cite{6dFGS}. 

Therefore, we illustrate two cases, namely, the results from the combination of WMAP, SPT, SNLS3 and BAO data as well as the results arising from the combination of WMAP, SPT, HST and BAO data.

\section{Current constraints}
\label{sec:results}
Table~\ref{tab:tab1} shows the $68\%$ and $95\%$ CL errors on the dark radiation parameters and on the size of the dark radiation-dark matter interaction strength arising from the two possible combinations of data sets considered here for both interacting and non interacting scenarios. Notice, first, that the $1-2\sigma$ preference found in the literature for extra dark radiation species is still present in both interacting and non interacting scenarios in which the dark radiation clustering properties are not standard. Overall, the bounds on $\Delta\neff$ are not largely modified when allowing for a dark radiation-dark matter coupling, see also the results presented in Ref.~\cite{concha}. However, the bounds on the dark radiation clustering properties $\ceff$ and $\cvis$ in the $\Lambda$CDM scenario and in minimal extensions of this scheme presented in Refs.~\cite{darkr,darkr2} are drastically changed when considering the possibility of an interaction between the dark radiation and the dark matter fluids. For instance, in Ref.~\cite{darkr}, in the context of a $\Lambda$CDM scenario,  it is found that $\ceff=0.24^{+0.08}_{-0.13}$ at $95\%$ CL. Similar results were found in Ref.~\cite{darkr2}, where the $\Lambda$CDM scenario was extended to consider other cosmological models with a dark energy equation of state or with a running spectral index. Indeed, within non interacting scenarios, we find $\ceff=0.32^{+0.04}_{-0.03}$ and $\cvis=0.27^{+0.34}_{-0.22}$ at $95\%$~CL from the combination of WMAP, SPT, HST and BAO data sets. These bounds are much weaker when allowing for an interacting dark radiation component: the errors on $\ceff$ are degraded by an order of magnitude, while the errors on $\cvis$ increase by a factor of two. We find, for the same combination of data sets than the one quoted above, $\ceff=0.28^{+0.44}_{-0.28}$ and  $0.45^{+0.52}_{-0.45}$, both at $95\%$~CL. 

\begin{figure*}[!htb]
\begin{tabular}{c c}
\includegraphics*[width=8.5cm]{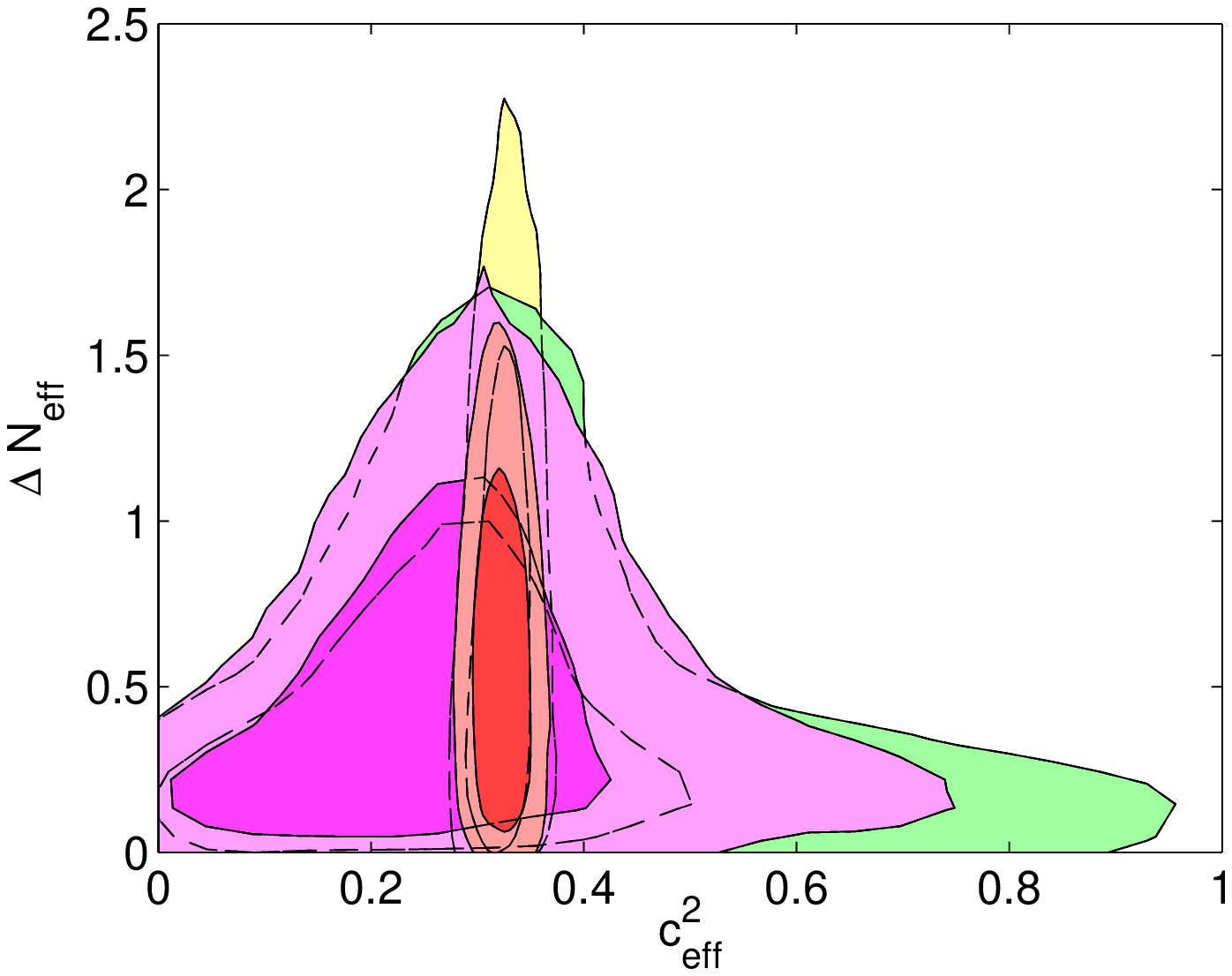}&
\includegraphics*[width=8.5cm]{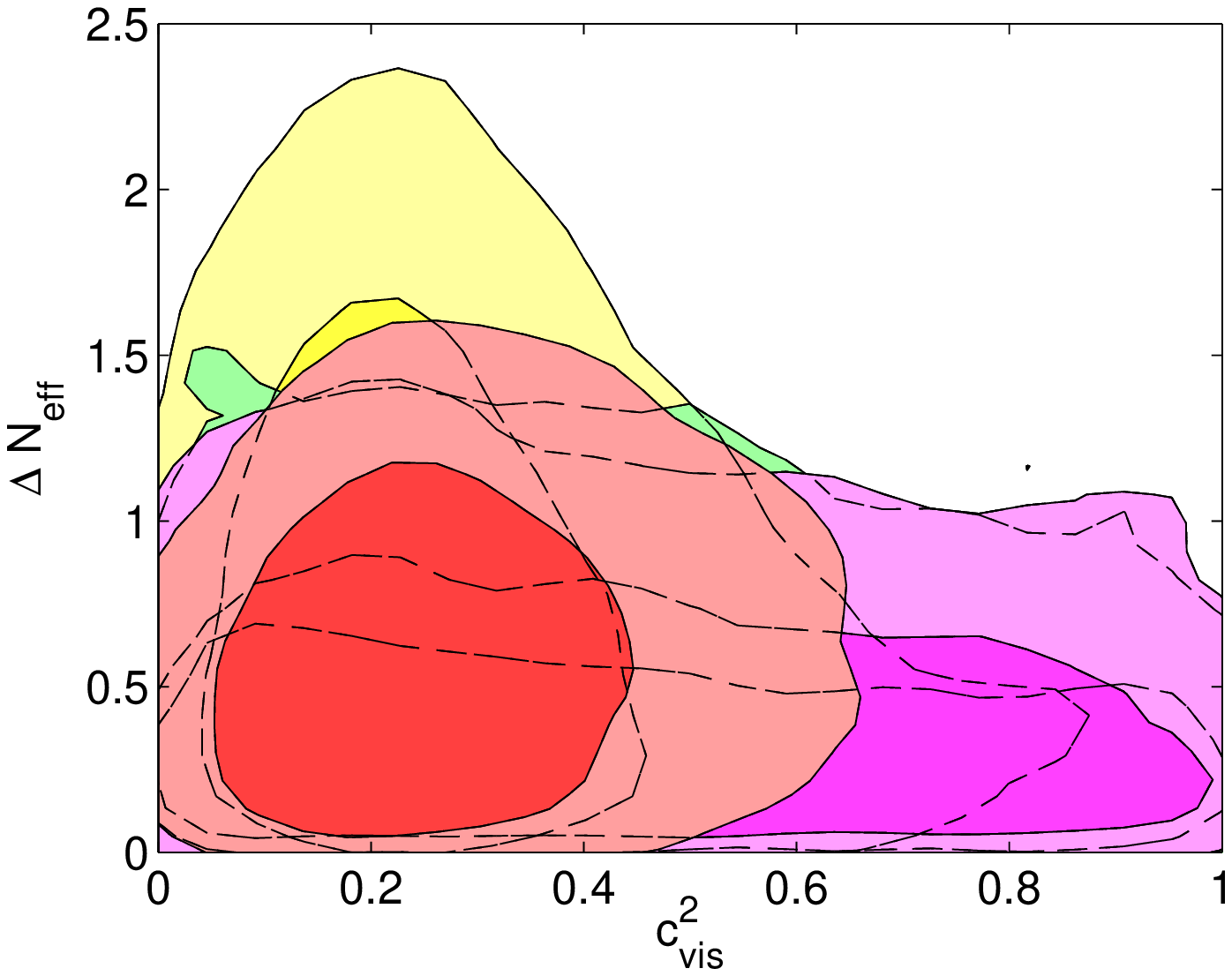}\\
\end{tabular}
\caption{\small{Left panel: $68\%$ and $95\%$ CL contours in the ($\ceff$, $\Delta\neff$) plane arising from the MCMC analysis of WMAP, SPT, BAO and HST/SNLS3 data. The green (yellow) contours refer to the case of  WMAP, SPT, BAO and SNLS3 data sets with (without) interaction between the dark radiation and dark matter fluids. The magenta (red) contours refer to the case of WMAP, SPT, BAO and HST data sets with (without) interaction between the dark radiation and dark matter sectors. Right panel: as in the left panel but in the ($\cvis$, $\Delta\neff$) plane.}}
\label{fig:current}
\end{figure*}

Figure~\ref{fig:current} (left panel) depicts the $68\%$ and $95\%$~CL allowed regions in the ($\ceff$, $\Delta\neff$) plane arising from the MCMC analysis of the cosmological data sets described in the previous section. We illustrate here the four cases shown in Tab.~\ref{tab:tab1}. The green (yellow) contours refer to the case of  WMAP, SPT, BAO and SNLS3 data sets with (without) interaction between the dark radiation and dark matter fluids. The magenta (red) contours refer to the case of WMAP, SPT, BAO and HST data sets with (without) interaction. Notice that the errors on the $\ceff$ parameter are largely increased when the interaction term is switched on, while the errors on $\Delta\neff$ are mildly affected by the presence of such an interaction. Notice that HST data is more powerful than SNLS3 data in constraining $\Delta\neff$, agreeing with previous results in the literature, see Ref.~\cite{Giusarma:2012ph}. The reason is because $\Delta\neff$ is highly degenerate with $H_0$, and HST data provide a strong prior on the former parameter.  

The right panel of Fig.~\ref{fig:current} depicts the $68\%$ and $95\%$~CL allowed regions in the ($\cvis$, $\Delta\neff$) plane, being the color code identical to the one used in the left panel. While the impact of the coupling is not as large as in the effective velocity case, the errors on the viscosity parameter $\cvis$ are enlarged by a factor of two in interacting dark radiation models.

\begin{table*}
\begin{center}
\begin{tabular}{lccccccc}
\hline \hline
 &            & WMAP+SPT+BAO$_{2012}$ &WMAP+SPT+BAO$_{2012}$ & WMAP+SPT+BAO$_{2012}$ &  WMAP+SPT+BAO$_{2012}$\\

 &             & +HST int. & +HST no int. & +SNLS3 int. &  +SNLS3 no int. \\
\hline
\hspace{1mm}\\
$\ceff$ &   &$0.28^{+0.10 +0.44}_{-0.12 -0.28}$ & $0.32^{+0.02 +0.04}_{-0.02 -0.03}$& $0.30^{+0.12 +0.50}_{-0.15 -0.30}$ & $0.32^{+0.02 +0.04}_{-0.02 -0.04}$ \\
\hspace{1mm}\\
$\cvis$ & &  $0.45^{+0.34 +0.52}_{-0.31 -0.45}$ & $0.27^{+0.13 +0.34}_{-0.13 -0.22}$ & $0.46^{+0.36 +0.51}_{-0.32 -0.46}$ & $0.27^{+0.13 +0.42}_{-0.14 -0.23}$\\
\hspace{1mm}\\
${\Delta \neff}$ & 68\%CL  & $ <0.81$ & $0.62^{+0.36 +0.80}_{-0.36 -0.53}$ & $<0.76$ & $0.77^{+0.50 +1.29}_{-0.72 -0.72}$ \\
                      & 95\%CL &  $<1.30$ & & $<1.47$ &  \\
\\

$Q_0$ & 68\%CL  & $ <0.8$ & --- & $<0.8$ & ---\\
    ($10^{-33}\textrm{cm}^2/ \textrm{MeV}^{-1}$) & 95\%CL &  $<4.9$ &--- & $<5.4$ & --- \\
\\
\hline
\hline
\end{tabular}
\caption{\small{1D marginalized bounds on the dark radiation parameters and on the size of the dark radiation dark matter interaction $Q_0$ using WMAP, SPT, BAO data and HST/SNLS3 measurements, see text for details. We show the constraints for both interacting and non interacting models, presenting the mean as well as the $68\%$ and $95\%$~CL errors of the posterior distribution.}}
\label{tab:tab1}
\end{center}
\end{table*}

\section{Forecasts from future cosmological data}
\label{sec:forecasts}
We evaluate here the constraints on the dark radiation parameters, $\Delta\neff$, $\ceff$, $\cvis$, by means of an analysis of future mock CMB data for the ongoing Planck experiment and the future COrE mission. These CMB mock data sets are then fitted using a MCMC analysis to a non interacting cosmological scenario but allowing the dark radiation parameters to have non standard values. The CMB mock data sets are generated accordingly to noise properties consistent with the Planck and COrE CMB missions. The fiducial $C_\ell$ model we use is a $\Lambda$CDM scenario (i.e. a flat universe with a cosmological constant and three massless active neutrino species) adding an interaction between the dark radiation and dark matter sectors with $Q_0=10^{-32}$cm$^2$ MeV$^{-1}$, assuming one dark radiation interacting species $\Delta\neff$=1 and standard clustering and viscosity parameters for the dark radiation, i.e. $\cvis=\ceff=1/3$. For each frequency channel we consider a detector noise given by $\omega^{-1}=\left(\theta\sigma\right)^2$, being $\theta$ the FWHM of the gaussian beam and $\sigma=\Delta T/T$ the temperature sensitivity (the polarization sensitivity is $\Delta E/E= \sqrt{2}\Delta T/T$). Consequently the $C_\ell$ fiducial spectra get a noise contribution which reads $N_\ell=\omega^{-1}\exp\left(\ell(\ell+1)/\ell^2_b\right)$, where $\ell_b\equiv\sqrt{8 ln 2}/\theta$.

\begin{figure*}[!htb]
\begin{tabular}{c c}
\includegraphics*[width=8.5cm]{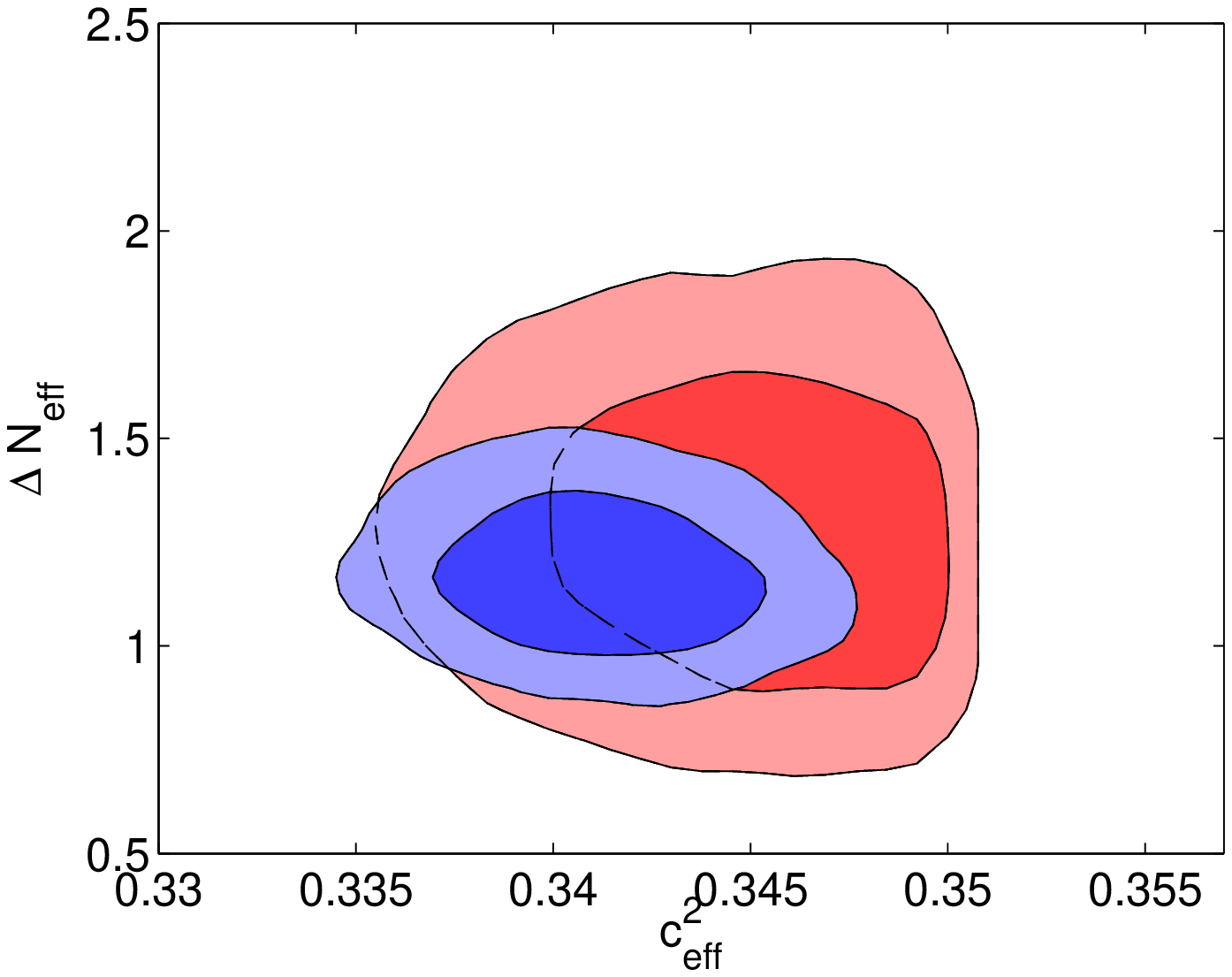}&
\includegraphics*[width=8.5cm]{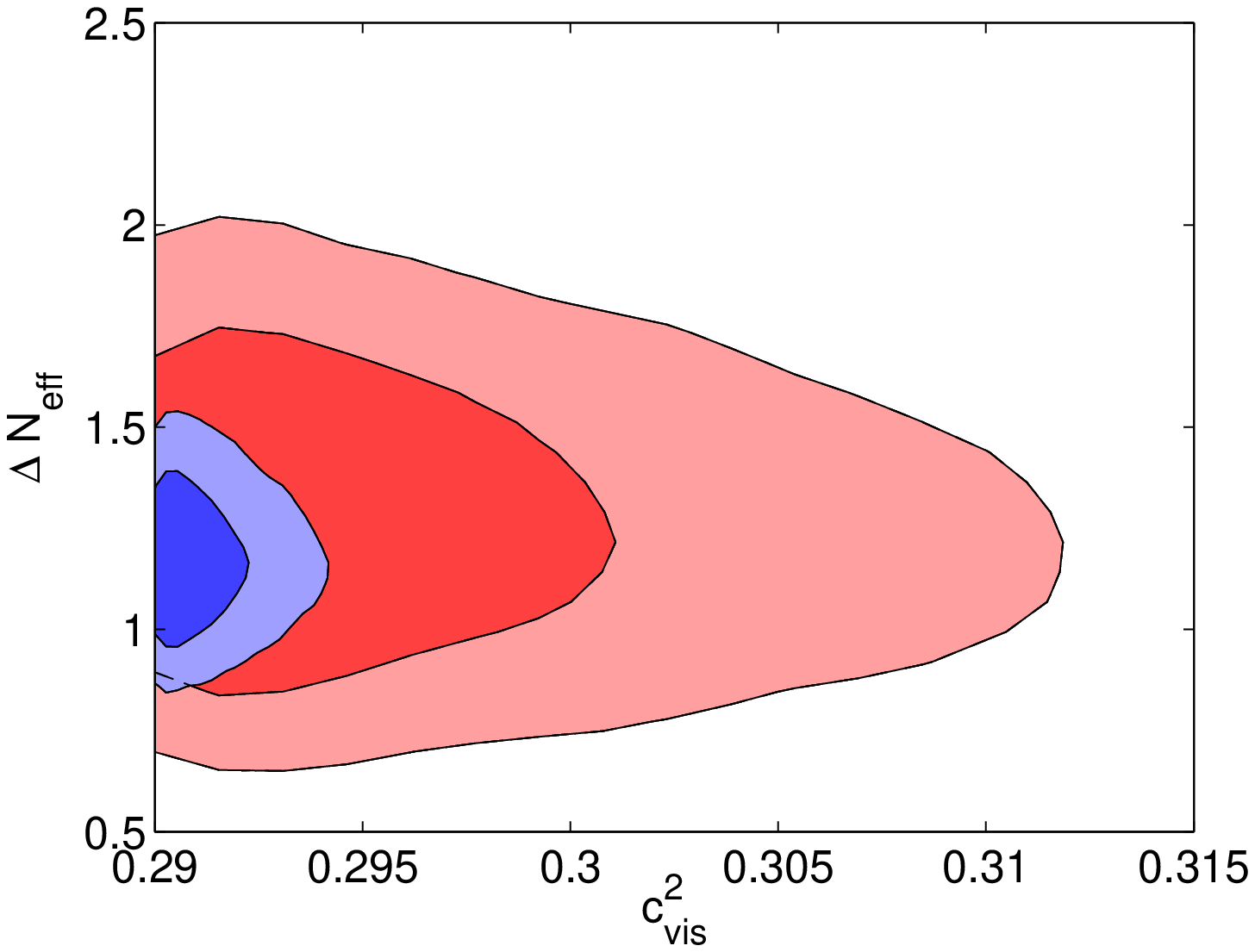}\\
\end{tabular}
 \caption{\small{Left panel: $68\%$ and $95\%$ CL contours in the ($\ceff$, $\Delta\neff$) arising from the MCMC analysis of Planck (red contours) and COrE (blue contours) CMB mock data. The mock data are generated adding an interaction between the dark radiation and dark matter sectors with $Q_0=10^{-32}$cm$^2$ MeV$^{-1}$, assuming one dark radiation interacting species $\Delta\neff$=1 and standard clustering and viscosity parameters for the dark radiation. The CMB mock data is then fitted to a non interacting cosmology but allowing the dark radiation parameters $\ceff$ and $\cvis$ to have non standard values. Right panel: as in the left panel but in the ($\cvis$, $\Delta\neff$) plane.}}
\label{fig:forecast}
\end{figure*}

Figure~\ref{fig:forecast} (left panel) depicts the $68\%$ and $95\%$ CL contours in the ($\ceff$,$\Delta\neff$) plane arising from the MCMC analysis of Planck and COrE mock data. Notice that the reconstructed value for $\ceff$ is larger than the simulated value $\ceff=1/3$. The reason for that is due to the degeneracy between the dark radiation-dark matter interaction $Q_0$ and $\ceff$, see Fig.~\ref{WS}, from which one would expect a negative correlation between the interaction cross section and the effective velocity. If such an interaction occurs in nature but future CMB data is analysed assuming a non interacting model, the reconstructed value of $\ceff$ will be higher than the standard expectation of $1/3$, see Tab.~\ref{tab:forecasts}. From what regards to $\cvis$, see Fig.~\ref{fig:forecast} (right panel), the effect is the opposite since these two parameters are positively correlated and therefore the reconstructed value of $\cvis$ is lower than the canonical $1/3$, see Tab.~\ref{tab:forecasts}. Therefore, if the dark radiation and dark matter sectors interact, a large bias on the dark radiation clustering parameters could be induced if future CMB data are analysed neglecting such coupling. On the other hand, the bias induced in $\Delta\neff$ is not very significant, being the reconstructed value consistent with the $\Delta\neff=1$ simulated one within $1\sigma$. 

\begin{table*}
\begin{center}\footnotesize
\begin{tabular}{lccc}
\hline \hline
  &            & Planck & COrE \\
\hline
\hspace{1mm}\\
$\ceff$ &   &$0.35^{+0.003+0.005}_{-0.003 -0.007}$ & $0.34^{+0.004 +0.006}_{-0.001 -0.003}$ \\
\hspace{1mm}\\
$\cvis$ & &  $0.30^{+0.005 +0.017}_{-0.006 -0.006}$ & $0.29^{+0.001 +0.002}_{-0.001 -0.001}$ \\
\hspace{1mm}\\
${\Delta\neff}$  & & $1.26^{+0.27+0.54}_{-0.25 -0.51}$ & $1.17^{+0.13 +0.27}_{-0.12 -0.23}$ \\
\hspace{0.5mm}\\
\hline
\hline
\end{tabular}
\caption{\small{Constraints on the dark radiation clustering parameters from the Plank and COrE mock data sets described in the text. We present the mean as well as the $68\%$ and $95\%$~CL errors of the posterior distribution. We have set $Q_0=10^{-32}$cm$^2$ MeV$^{-1}$, $\ceff=\cvis=1/3$ in the mock data sets. Then, we have fitted these data to non interacting models in which both $\ceff$ and $\cvis$ are free parameters.}}
\label{tab:forecasts}
\end{center}
\end{table*}

%%%%%%%%%%%%%%%%%%%%%%%%%%%%%%%%%%%%%%%%

\section{Conclusions}
\label{sec:conclusions}
Standard dark radiation is made of three light active neutrinos. However, many extensions of the standard model of elementary particles predict an extra dark radiation component in the form of sterile neutrinos, axions or other very light degrees of freedom which may interact with the dark matter sector. In fact, once that one assumes the existence of extra dark radiation species as well as the existence of a dark matter sector there is a priori no fundamental symmetry which forbids couplings between these two dark fluids. If one allows for such a possibility, the clustering properties of these extra dark radiation particles might not be identical to those of the standard model neutrinos, since the extra dark radiation particles are coupled to the dark matter. In this paper we have analyzed the constraints from recent cosmological data on the dark radiation abundances, effective velocities and viscosity parameters.
 While the bounds on $\Delta\neff$ are very close to those of uncoupled models, the errors on the clustering dark radiation properties are largely increased, mostly due to the existing degeneracies among the dark radiation-dark matter coupling and $\ceff$, $\cvis$. The cosmological bounds on the dark radiation effective velocity $\ceff$ found in non-interacting schemes are degraded by an order of magnitude when a dark radiation-dark matter interaction is switched on. In the case of the viscosity parameter $\cvis$, the errors on this parameter are a factor of two larger when considering interacting scenarios. We have also explored the perspectives from future Cosmic Microwave Background data. If dark radiation and dark matter interact in nature, but the data are analysed assuming the standard, non interacting picture, the reconstructed values for the effective velocity and for the viscosity parameter will be shifted from their standard $1/3$ expectation, namely $\ceff=0.34^{+0.006}_{-0.003}$ and $\cvis=0.29^{+0.002}_{-0.001}$ at $95\%$~CL for the future COrE CMB mission.

\section{Acknowledgments}
O.M. is supported by the Consolider Ingenio project CSD2007-00060, by PROMETEO/2009/116, by the Spanish Ministry Science project FPA2011-29678 and by the ITN Invisibles PITN-GA-2011-289442.
%%%%%%%%%%%%%%%%%%%%%%%%%%%%%%%%%%%%%%%%%%%

%%%%%%%%%%%%%%%%%%%%%%%%%%%%%%%%%%%%%%%%
%%%%%%%%%%%%%%%%%%%%%%%%%%%%%%%%%%%%%%%%

\end{document}